# Optimized Fabrication Procedure for High-Quality Graphene-based Moiré Superlattice Devices

Shuwen Sun[1] and Pablo Jarillo-Herrero[1]

*Department of Physics, Massachusetts Institute of Technology, Cambridge, Massachusetts, 02139, USA.*

Correspondence to: Pablo Jarillo-Herrero at pjarillo@mit.edu

**SUMMARY:**
This article presents an optimized, experience-informed protocol for fabricating high-quality graphene-based moiré superlattice devices with precise twist angles, utilizing a modified dry transfer technique based on a highly tunable, custom-built transfer setup.

**ABSTRACT:**
Moiré superlattices constitute a versatile platform to investigate emergent phenomena arising from the interplay of strong correlations and topology, while offering flexible in situ tunability. However, the fabrication of such moiré superlattices is challenging. It is difficult to achieve highly uniform devices with a precise twist angle because of the unintentional introduction of heterostrain, twist angle disorder, and angle/lattice relaxation during the nanofabrication process. This article introduces an optimized, experience-informed protocol for fabricating high-quality graphene-based moiré superlattice devices, focusing on a modified dry transfer technique. The transfer process is performed in a highly tunable, custom-built transfer setup that enables precise position, angle, and temperature control. By combining rigorous flake selection criteria, pre-cleaned bubble-free bottom gates, and graphene laser ablation, the moiré superlattice is constructed by deliberately overlaying twisted graphene flakes at a submicron speed at room temperature. Through precise control of the transfer process, the resulting graphene moiré superlattice devices exhibit high uniformity and desired twist angles. This optimized protocol addresses existing challenges in the fabrication of graphene-based moiré superlattice devices and paves the way for further advances in the rapidly evolving field of moiré materials.

**INTRODUCTION:**
A key driving force of modern condensed matter physics is the combination of strong correlations and topology into a single tunable physical system. Since the discovery of correlated insulator states[1] and superconductivity[2] in magic-angle twisted bilayer graphene (MATBG), its remarkable properties have stimulated a growing number of studies on moiré materials[3]. In recent years, a variety of graphene-based moiré superlattices have been constructed and investigated. This rapidly growing field has revealed a wealth of emergent phenomena, including but not limited to exotic correlated insulators[4–9], unconventional superconductivity[10–14], nematicity[15], and orbital ferromagnetism[16–18].

The rich physics unveiled in these graphene moiré superlattices benefits from the unprecedented *in situ* tunability offered by the small twist angles between graphene layers. Meanwhile, it is also impacted by the uncertainties arising from this additional twist angle degree of freedom. The twist angle is created by overlaying two graphene flakes with a relative rotation of their crystal axes. To achieve the magic-angle regime (1°~1.2°), where correlations play a dominant role, it is imperative to precisely control the twist angle with sub-0.1° accuracy. However, during the fabrication procedure, various detrimental factors can be introduced, such as heterostrain[19, 20], twist angle disorder[21–23], and lattice relaxation[24, 25], among others. Using imperfect flakes or failing to control the pickup speed of graphene can result in excessive strain and angular disorder. As the small twist angle configuration is not the thermodynamic ground state, incorporating unnecessary high-temperature transfer processes increases the risk of undesired lattice relaxation. These factors reduce the success rate of graphene moiré superlattice device fabrication and are especially prominent in the magic-angle regime. They pose significant experimental challenges[26] for researchers in this field, making it difficult to obtain devices with the desired twist angle or even reproducing previous measurement results.

In this article, with the objective of alleviating the above challenges, an optimized, experience-informed, and step-by-step protocol for fabricating high-quality graphene-based moiré superlattice devices is presented. Based on a modified dry transfer technique[27, 28], the transfer process is performed using a highly tunable, custom-built transfer setup with precise position, angle, and temperature control. At the heart of this protocol are rigorous flake selection criteria, pre-cleaned bubble-free bottom gates, graphene laser ablation, and, most importantly, smooth pickup of the pre-cut graphene flakes at a submicron speed at room temperature. Apart from these crucial aspects, minor factors that possibly affect the experimental outcomes are also discussed. Even though the presented protocol uses twisted bilayer graphene as an example, it is applicable to all kinds of graphene-based moiré superlattice structures. The detailed workflow aims to enhance accessibility to the fabrication of high-quality graphene moiré superlattice devices for the broader research community.

**PROTOCOL:**

**The video demonstration of the PROTOCOL can be accessed via the article page on the publisher's website.** The materials and instruments used throughout the protocol can be found in the **Table of Materials.**

**1    Preparation of two-dimensional (2D) material flakes**

**1.1    Graphene exfoliation**

1.1.1    Cut a $Si/SiO_2$ wafer into 2 cm × 2 cm pieces using a tungsten carbide scriber. Clean the wafer surface afterward with a nitrogen spray gun to remove silicon dust.

Note: Wafers with 285 nm thick $SiO_2$ are typically used for graphene and hBN exfoliation, as this thickness provides a good color contrast for visual identification of 2D materials.

1.1.2 Preheat the hot plate to 250 °C, then place the Si/SiO$_2$ wafers on the hot plate.

1.1.3 Place a graphite crystal on a rectangular silicone-free blue tape (1005R). Fold the tape in half to cover the crystal. Press the graphite crystal firmly, then gently unfold the tape. It peels off the top graphite layers, exposing a flat and shiny crystal surface.

1.1.4 Place another rectangular blue tape onto the freshly exfoliated graphite surface. Gently press it, then peel it off. It transfers the top layers of fresh graphite onto the new tape.

1.1.5 Inspect the transferred graphite layers. The quantity of crystal is crucial for optimal exfoliation results. If needed, repeat the procedure in 1.1.4 until a sufficient amount of crystal is accumulated on the new tape (Fig. 1A).

1.1.6 Gently fold and unfold the new tape multiple times, distributing the graphite across the tape (Fig. 1B). Stop when the surface is uniformly covered with graphite thin layers.

Note: Folding the tape too many times may reduce the yield of large graphene flakes.

1.1.7 Examine the new tape against a light source (Fig. 1C). Ideally, the tape is covered with transparent graphite islands. If it has too many thick graphite chunks, place another blue tape on top. Peeling it off will help reduce the graphite thickness.

1.1.8 Cut the prepared blue tape into pieces sized to match the Si/SiO$_2$ wafers. Remove the wafers from the hot plate and immediately apply the tape.

1.1.9 Allow the wafers to cool down to room temperature, then gently peel off the tape. Inspect the exfoliation results using an optical microscope. If excessive tape residue is present on the wafer, consider lowering the heating temperature for the next attempt.

1.2 **Hexagonal boron nitride (hBN) exfoliation**

1.2.1 Cut and clean the wafers as described in 1.1.1. Preheat the hot plate to 160°C.

1.2.2 Place 2–3 shiny hBN crystals on a rectangular silicone-free blue tape (1009R) (Fig. 1D).

Note: Different tapes are used for graphene and hBN exfoliation. The 1005R tape has stronger adhesion than the 1009R, resulting in higher flake densities but more tape residue on the wafers.

1.2.3 Gently fold and unfold the tape several times to distribute the hBN crystals across the tape surface (Fig. 1E). Stop when the tape is covered with thin and shiny hBN crystals.

1.2.4 Inspect the tape surface under reflected light (Fig. 1F). If there are too many thick crystals, use another blue tape to reduce their thickness, as described in 1.1.7.

1.2.5 Cut the prepared blue tape into pieces matching the size of the Si/SiO$_2$ wafers. Smoothly apply the tape onto the wafers at room temperature.

1.2.6 Place the taped wafers on the hot plate at 160 °C and heat them for 2 minutes. After heating, once the wafers cool to room temperature, gently peel off the tape at a slow and controlled speed of approximately 2 minutes per wafer.

1.2.7 After removing the tape, place the wafers in a quartz tube inside a furnace and anneal at 350 °C for 10 hours in a forming gas atmosphere (50% $H_2$ – 50% Ar). This step aims to reduce tape residue on the wafer surface.

Note: Other annealing conditions, such as high vacuum, can achieve similar results.

1.2.8 After the annealing process is complete, remove the wafers from the furnace and use a nitrogen spray gun to blow off any glass particles introduced during the annealing.

1.2.9 Anneal the quartz tube at 1000 °C for 1 hour while exposing it to the atmosphere. This step removes chemical residues from the tube in preparation for the future annealing process.

1.3 **Flake selection of 2D materials**

1.3.1 For 2D material flake selection, use an optical microscope with a 5x objective to quickly prescreen the wafers under bright-field illumination. Identify flakes with the desired size, shape, and thickness according to the flake selection criteria specified in the DISCUSSION section. Representative flakes that meet the criteria are shown in Fig. 1G - I.

1.3.2 Use a 50x or 100x objective to assess flake quality by adjusting the color contrast of the bright-field image. Alternatively, use dark-field illumination with sufficient exposure time to check the flake. Both strategies help further eliminate imperfect flakes.

1.3.3 To verify the flake quality, consider performing atomic force microscopy (AFM) in tapping mode to check the flake surface topography. Only select flakes with a clean surface and no apparent strain for further processing.

2 **Fabrication of glass slides with PC/PDMS stamps**

This section describes the fabrication process of glass slides with polycarbonate (PC) and polydimethylsiloxane (PDMS) stamps, which are used for picking up 2D materials.

2.1.1 Prepare a clean 6'' petri dish. Mix the silicone elastomer base with the curing agent in the petri dish, with a mixing weight ratio of 10:1. To get an optimal PDMS thickness of around 1-2 mm, a mixture of around 10 g base and 1 g curing agent is usually used.

2.1.2 Once fully mixed, place the petri dish on a flat surface and cure it for two days at room temperature. After curing, the PDMS is ready to use.

2.1.3 Attach a double-sided tape with a punched hole to one end of a clean glass slide (Fig. 2A). Then, position a clean PDMS stamp (1–2 mm thick) at the center of the hole (Fig. 2B).

Note: To achieve strong adhesion between the PDMS stamp and the glass slide, the interface must be clean, uniform, and free of dust or bubbles.

2.1.4 In the fume hood, take another clean glass slide and dispense an appropriate amount of 6% w/w PC in chloroform solution onto the surface (as shown in Fig. 2C).

2.1.5 Place a clean glass slide on top of the bottom slide and gently press it (Fig. 2D). Once the solution is spread, slide the two glass slides past each other to create a smooth PC film. After the chloroform evaporates, a PC thin film forms on the bottom slide (Fig. 2E).

2.1.6 Place another piece of double-sided tape with a punched hole over a clean and uniform region of the PC film. Press around the hole using the back of a tweezer to ensure a secure attachment between the tape and the PC film (Fig. 2F).

2.1.7 Cut the PC film along the edges of the tape to separate this section from the remaining part on the glass slide, then gently peel off the tape (Fig. 2G).

2.1.8 After peeling off the tape, inspect the free-standing PC film at the center of the hole (Fig. 2H). It is preferable to prepare the PC film in a high-humidity environment, such as on rainy days. Otherwise, the film may wrinkle during the peeling process. If wrinkling occurs, repeat steps 2.1.6 to 2.1.7 until a uniform PC film is achieved.

2.1.9 Overlay the free-standing PC film onto the PDMS stamp prepared in step 2.1.3. Then use a razor blade to press the double-sided tape around the hole firmly together (Fig. 2I).

2.1.10 Cut away the excess tape beyond the central PC/PDMS stamp area. A glass slide with a PC/PDMS stamp on one end is fabricated (Fig. 2J), with the zoom-in view shown in Fig. 2K.

## 3 Dry transfer of graphene moiré superlattice (twisted bilayer graphene as an example)

Note: For details on the custom-built transfer setup (Fig. 3A) and the integrated supercontinuum laser (Fig. 4A) used in this protocol, please refer to the DISCUSSION section.

### 3.1 Graphene laser ablation

3.1.1 Place the graphene wafer on the sample stage and locate the graphene using the 10x objective. Adjust the flake to the desired orientation, then switch to the 50x objective.

3.1.2 Use the "Draw Bezier curves and straight lines" function in *Inkscape* to outline the graphene boundary and the planned laser-cutting lines. If the graphene has regions with minor folds or cracks, use laser-cutting lines to separate them from the core region. It can reduce the risk of further folding or cracking during the graphene pickup process.

Note: The laser cutting lines can divide the graphene into two or more pieces, depending on the intended structure, with each piece measuring at least 20 μm in width and 30 μm in length.

3.1.3 Power on the laser and increase its intensity until the beam spot becomes visible. Adjust the sample stage to align the beam spot with the start of a planned laser-cutting line. Then, adjust the laser intensity to a suitable level for graphene ablation.

Note: The laser power required to cut graphene without damaging the $SiO_2$ surface is between 60 mW and 150 mW, as measured before the beam splitter.

3.1.4 Move the sample stage to guide the beam spot along the planned laser-cutting line to cut through the flake. After cutting, zero the laser intensity and inspect the laser-cutting line.

3.1.5 Repeat the procedure to make an additional cut adjacent to the existing one. This increases the gap between the graphene pieces, providing more space for the hBN straight edge to align without contacting the neighboring graphene piece.

3.1.6 Repeat the procedure for each planned laser-cutting line. After completing the laser ablation, the pristine graphene is divided into several pieces (Fig. 4B). The central pieces used for building the moiré superlattice are isolated from the rest of the flake.

3.1.7 Remove the graphene wafer from the stage and inspect the laser-cutting lines using an optical microscope. Ensure that the cutting lines are free of any contaminants.

Note: The graphene laser ablation technique can also be used to engineer graphite flakes into desired geometries. For instance, an irregular graphite flake (Fig. 4C) can be shaped into a perfect rectangular graphite finger (Fig. 4D), which is ideal for use as a gate electrode. This approach significantly reduces the time spent searching for graphite flakes with specific geometries.

3.2 **Preparation of graphite bottom gates**

3.2.1 Pick up the bottom hBN flake (Fig. 5A - D) following similar steps in section 3.3. Notably, for the bottom hBN flake, a straight crystal edge and a parallel wavefront are not required (for the definition of the wavefront, see 3.3.8).

3.2.2 Place the wafer with the graphite finger onto the sample stage. Center the graphite finger in the camera window and orient it in the desired direction. Use *Peek Through* to make the *Inkscape* window semi-transparent while floating above the camera window. Outline the boundary of the graphite finger in *Inkscape*.

3.2.3 Set the sample stage temperature to 80 °C. Load the glass slide and move it to the pre-engage position (See step 3.3.4 and Fig. 3B).

3.2.4 Following similar steps to those described in 3.4.5 - 3.4.10, align the bottom hBN flake and the graphite finger before the stamp contacts the wafer, as shown in Fig. 5E. This alignment ensures that the graphite finger is positioned underneath the clean and uniform region of the bottom hBN flake.

3.2.5 Use the Z direction actuator to engage the PC/PDMS stamp onto the wafer at a speed of 5 µm/s until the wavefront is close to the graphite finger (Fig. 5F).

3.2.6 Reduce the speed to 0.5 µm/s to allow the wavefront to slowly pass the graphite finger until it has completely covered the graphite (Fig. 5G).

3.2.7 Disengage the wavefront at a speed of 0.5 µm/s. The color of the graphite finger will change from purple to semitransparent dark grey once picked up (Fig. 5H). When the graphite is fully picked up, completely disengage the wavefront at a speed of 5 µm/s.

3.2.8 Hot-release the bottom hBN flake and the graphite gate onto a pre-patterned marker wafer at a sample stage temperature of 160°C, contacting the Ti/PdAu metal finger (Fig. 5I - L), following similar steps as described in section 3.5.

3.2.9 Remove the PC film by immersing the wafer in chloroform for an hour. After removal, rinse the wafer with isopropyl alcohol and blow dry with a nitrogen spray gun.

Note: Normally, 20 minutes in chloroform is sufficient to dissolve the PC film. Here, one hour is used to achieve a cleaner gate surface.

3.2.10 Place the wafer in a quartz tube and anneal at 350 °C for 10 hours in a forming gas atmosphere (50% $H_2$ – 50% Ar).

3.2.11 Inspect the bottom gate using the optical microscope. The bright-field and dark-field microscopic images of a bubble-free bottom graphite gate are shown in Fig. 6A, B.

3.2.12 Use AFM tapping mode to scan the bottom gate. Select the gate regions free of bubbles and non-uniform structures, with the width of the region slightly larger than that of the gate. The gate surface after annealing is covered with polymer residue, shown in Fig. 6C.

3.2.13 After selecting the uniform and bubble-free gate regions, switch the AFM to the standard contact mode with the same tapping mode tip used in the last step.

Note: The tapping mode tip is stiffer than the contact mode tip, giving better cleaning results.

3.2.14 Scan the selected regions with a resolution of 512 lines per scan and an applied normal force on the tip of a few tens of nN at a speed of around 20 µm/s to clean away the polymer residue on the surface. This step is referred to as the tip-cleaning process.

3.2.15 After one round of tip-cleaning, switch the AFM back to the tapping mode and check the cleaning results with a new tapping mode tip.

3.2.16 Repeat the tip-cleaning process until the polymer residue is fully removed from the gate region to the sides (Fig. 6D).

3.3 **Pick up the top hBN flake**

3.3.1 Place the wafer on the sample stage and locate the top hBN flake using the 10x objective.

3.3.2 Adjust the orientation of the hBN flake until its straight edge is vertical and positioned on the right side of the flake. Then, outline the flake boundary in *Inkscape*. This drawing serves as a reference for aligning the hBN to the graphene flakes later.

Note**:** The zoom levels of the camera and *Inkscape* must be kept consistent throughout the entire transfer process to maintain a uniform scale between the flakes and the drawings.

3.3.3 Gently clamp the glass slide with the PC/PDMS stamp inside the socket at a downward tilt angle of 2° - 3°. Slide the socket to the left using the sliding tray until the glass slide is positioned above the sample wafer, then manually lock it in place.

3.3.4 Set the sample stage temperature to 50 °C. Engage the glass slide downward using the Z direction actuator with a fast speed of 1 mm/s, until the stamp is 2 mm above the wafer. Now, the system is in the pre-engage position (Fig. 3B).

3.3.5 Focus the microscope on the PC film and inspect the surface cleanness (similar to Fig. 5A). Select a clean region slightly left of the stamp center for contact with the hBN flake, as the stamp will engage from the left. Move the hBN flake to the selected clean region.

3.3.6 Engage the glass slide further downward at a speed of 0.1 mm/s. When the PC film and the wafer are nearly in the same focal plane, reduce the speed to 5 µm/s.

3.3.7 Identify the touching point by observing Newton's rings formed in between the stamp and the wafer. To facilitate a smooth pickup process, ensure that the hBN is positioned to the right and away from the touching point. If this is not the case, return to the pre-engage position, and adjust the hBN location by repeating steps 3.3.5 to 3.3.7.

3.3.8 When the stamp contacts the wafer, their interface exhibits a green touching region, with the colorful right edge referred to as the wavefront. A parallel wavefront, as shown in Fig. 7A, is easier to be pinned at the hBN straight edge.

Note: Due to the potential tilted stamp surface, the wavefront may have a slight tilt compared to the hBN straight edge. To ensure they are parallel while picking up the graphene, record the approximate tilting angle and remember to compensate for it using the goniometers in step 3.4.3.

3.3.9 Further engage the stamp onto the wafer with a speed of 5 µm/s, until it completely covers the hBN flake (similar to Fig. 5C). Then, set the sample stage temperature to 80 °C.

Note: If the adhesion level of the PC film is uncertain, initially set the stage temperature to 60 °C, then gradually increase it until the hBN can be picked up smoothly.

3.3.10 After the temperature is reached, disengage the stamp with 5 µm/s until the wavefront is close to the hBN straight edge. Then, reduce the speed to 2 µm/s to slowly pick up the hBN. Once picked up, increase the speed to 5 µm/s to detach the PC film from the wafer.

Note: When the retracting wavefront touches the hBN straight edge, carefully monitor the color of the flake. It becomes transparent upon successful pickup. If the flake remains colorful, reengage the stamp to cover the flake and repeat the above step at a higher temperature. The PC film will become more adhesive to both the flake and the wafer as the temperature rises. The recommended upper temperature limit is 120 °C. Exceeding 120 °C may cause the PC film to detach from the PDMS stamp during disengaging. If the PC film is not sticky enough, consider applying UV/Ozone treatment to the glass slide with a stamp for 5-10 minutes. This treatment improves the adhesion of the PC film to 2D materials, increasing the success rate of flake pickup.

3.3.11 Switch off the stage heater and open the water-cooling system. Wait until the temperature drops below 45 °C, then close the water-cooling system and disengage the glass slide all the way up with 1 mm/s speed. Take off the hBN wafer from the stage.

Note: Moving the glass slide drastically at high temperatures disturbs fragile flakes, potentially causing folds and cracks to appear after cooling down. If severe folds or cracks develop, the flake should be discarded, as these uneven structures can hinder the smooth pickup of graphene.

### 3.4 Pick up the graphene moiré superlattice

3.4.1 Place the wafer with the freshly laser-cut graphene on the sample stage. Locate the graphene using the 10x objective, then adjust the flake orientation so that the laser cutting lines are parallel to the hBN straight edge, as shown in Fig. 7B.

3.4.2 Outline the graphene boundary in *Inkscape*, including the laser-cutting lines. Use *Peek Through* to make the drawing semi-transparent while floating above the camera window.

3.4.3 Adjust the goniometers to compensate for the tilt between the wavefront and the hBN straight edge, based on the observation in step 3.3.8.

3.4.4 Load the glass slide with the top hBN flake and move it to the pre-engage position, with the PC/PDMS stamp 2 mm above the graphene wafer.

3.4.5 Align the hBN and graphene drawings by positioning the hBN straight edge in the middle of the laser-cutting line that separates the central graphene pieces. Tune the microscope to focus on the graphene, then align it to its drawing in *Inkscape*.

3.4.6 Focus on the top hBN flake and move the glass slide in the XY directions, aligning the hBN to its drawing. This completes the rough alignment at the pre-engage position.

3.4.7 Focus on the pre-cut graphene, then adjust the focal plane slightly upward. Engage the glass slide downward using 0.1 mm/s speed until the top hBN flake comes into focus. At this point, the hBN and pre-cut graphene are in close proximity.

3.4.8 Refocus on the graphene. Set the speed to 5 μm/s and slowly engage the stamp until the hBN is roughly visible, ensuring the stamp and the wafer do not come into contact.

3.4.9 Realign both the graphene and top hBN to their drawings. This will serve as the final alignment before the stamp contacts the wafer. The flakes are well-aligned, as shown in Fig. 7C.

Note: Once the stamp contacts the wafer, further alignment adjustments should be avoided. Moving the stamp can induce excessive strain into the PC film and the hBN.

3.4.10 Continue to engage the stamp downward until it contacts the wafer. If the wavefront remains non-parallel to the hBN straight edge, restart from the step 3.4.3 to readjust the goniometers. The goal is to achieve a parallel wavefront, as seen in Fig. 7A.

3.4.11 Assess the hysteresis level of the z-axis stage by continuously moving the stamp upward while monitoring the wavefront movement. If it first moves forward for a few μm then starts retracting, it indicates that the stage exhibits obvious hysteresis with this glass slide. On the other hand, if the wavefront retracts immediately without any forward movement, the stage exhibits no obvious hysteresis with this glass slide.

Note: Hysteresis can impact the wavefront pinning at the hBN straight edge and lead to over-engaging, resulting in contamination of the subsequent graphene pieces. Therefore, if hysteresis is observed, the DC current heating method (step 3.4.13) is preferably used for the following engaging process instead of the mechanical method (step 3.4.12).

3.4.12 Mechanical method (without hysteresis): Use 5 μm/s speed to engage the stamp until the wavefront is close to the left edge of the hBN (Fig. 7D). Reduce the speed to 0.02 μm/s, then use this extremely slow speed to engage the hBN onto the first graphene piece (Fig. 7E). Once it completely covers the first graphene piece and gets pinned at the hBN straight edge (Fig. 7F), stop the engaging process. Clear the hysteresis by moving the stage upward with a speed of 5 μm/s until the wavefront tends to retract.

Note: Closely monitor the slow engaging process, as the wavefront may over-engage and contaminate the next graphene piece if it is not stopped promptly at the hBN straight edge.

3.4.13 DC current heating method (with hysteresis): Use 5 μm/s speed to engage the stamp until the wavefront is close to the left edge of the hBN (Fig. 7D). Clear the hysteresis by disengaging the stamp upward until the wavefront starts to retract. Apply a small DC current to the sample stage heater to heat up the stage slowly and controllably, with a temperature increasing speed of 0.5-1 °C/min. Due to the heat expansion of the stamp, the wavefront will engage continuously (Fig. 7E). Once it completely covers the first graphene piece and gets pinned at the hBN straight edge (Fig. 7F), stop the DC current.

3.4.14 Set a disengaging speed of 0.02 μm/s to pick up the graphene gently (Fig. 7G). Wait for the wavefront to move away from the hBN, then increase the speed to 5 μm/s to fully detach the stamp from the wafer.

3.4.15 Once the stamp and the wafer are fully detached, move the stamp upward for an additional 3 seconds at 5 μm/s speed to make sure all the flakes are detached from either the PC film or the wafer. Otherwise, once twisted, they will wrinkle the PC film.

3.4.16 Copy the entire graphene drawing in *Inkscape* and rotate it by the desired twist angle (±1.13° for MATBG). Change the copied drawing to a different color to prevent confusion (blue in Fig. 7H). Align the second graphene piece in the copied drawing with the first piece in the original drawing, maximizing their overlap.

Note: Based on experience, the final twist angle in the device is generally smaller than the angle targeted during the transfer process. Therefore, it is usually set to be slightly higher (~ 0.05°) than the "magic angle", which is the reason for targeting ±1.13° for MATBG device fabrication.

3.4.17 Rotate the sample stage by the desired twist angle (±1.13° for MATBG). Align the remaining graphene on the wafer to the copied graphene drawing. The precise twist angle control is achieved through a DC servo motor in the rotation stage, which provides 360° continuous rotation with a 2 μrad resolution and negligible hysteresis.

3.4.18 Focus on the top hBN flake on the glass slide and align it to the drawing. Given that the hBN and the remaining graphene are already in close proximity, this alignment functions as the final alignment (Fig. 7H).

3.4.19 Repeat steps 3.4.12 to 3.4.14 to pick up the twisted graphene piece. If this is the last graphene piece to pick up, the wavefront can over-engage across the hBN straight edge to squeeze out the bubbles inside the stack more completely.

3.4.20 At this stage, the graphene moiré superlattice is completely picked up by the top hBN flake, collectively referred to as the top stack.

3.4.21 Examine the quality of the top stack using an optical microscope. Save the color-contrasted bright-field image for use in step 3.5.2. In Fig. 7I, a high-quality twisted bilayer graphene top stack with few visible bubbles is presented.

3.5 **Encapsulate the graphene moiré superlattice**

3.5.1 Place the wafer with the pre-cleaned bubble-free bottom gate on the sample stage. Locate the bottom gate using the 10x objective.

3.5.2 Import the color-contrasted bright-field image of the top stack into *Inkscape* and align it to the drawings. Locate and outline the boundary of the bubble-free graphene moiré superlattice region of the top stack, which will be overlayed on the bottom gate.

3.5.3 Adjust the gate orientation to cover the largest area of the bubble-free top stack. Then, outline the boundary of the bottom gate.

3.5.4 To tear off the PC film after encapsulation, set the sample stage temperature to 160 °C, which is above the glass transition temperature of PC (~ 147 °C).

3.5.5 Load the glass slide with the top stack, then move it to the pre-engage position. Roughly align both the bottom gate and the top stack to the drawings.

3.5.6 Focus on the bottom gate, then slightly raise the focal plane. Engage the glass slide downward at a speed of 0.1 mm/s until the top stack becomes visible in the focal plane.

3.5.7 Identify another middle focal plane between the bottom gate and the top stack. Set the speed to 5 µm/s and gradually engage the stamp until the top stack comes into view.

Note: At a temperature above 120 °C, once the stamp contacts the wafer, it becomes very difficult to separate them. Using the middle focal plane as a reference ensures that the stamp and the wafer do not come into contact until the optimal alignment is achieved.

3.5.8 Repeat step 3.5.7 until the bottom gate and the top stack are almost in the same focal plane. Align them to the drawings accurately. This serves as the final alignment before the stamp contacts the wafer. Meanwhile, the sample stage reaches approximately 160 °C.

Note: High temperature causes air convection, leading to camera image oscillation, which complicates precise alignment. However, it is essential to complete the alignment quickly to prevent thermally induced twist angle relaxation. Therefore, it requires considerable practice.

3.5.9 Use a speed of 5 µm/s to engage the stamp until it touches the wafer. When the wavefront is close to the bottom gate, reduce the speed to 0.5 µm/s.

3.5.10 Slowly engage the top stack onto the bottom gate. Once the wavefront passes the top stack, switch back to 5 µm/s to engage the entire stamp onto the wafer (similar to Fig. 5J).

Note: Carefully control the wavefront when engaging the top stack. A smooth process increases the likelihood of maintaining the target twist angle in the final stack.

3.5.11 Wait for one minute to ensure that the PC film is fully softened, then disengage the stamp at a speed of 5 µm/s. The PC film remains adhered to the wafer while the PDMS stamp retracts, forming a transparent wavefront in between (similar to Fig. 5K).

3.5.12 As soon as the wavefront is fully retracted, the PC film and the PDMS stamp will be completely detached. Lift the stamp upward for an additional 3 seconds at a speed of 5 µm/s, then begin to move it around to tear off the PC film (similar to Fig. 5L).

3.5.13 After tearing off the PC film, fully disengage the glass slide at a speed of 1 mm/s, then take it off from the socket. Switch off the stage heater and activate the water-cooling system. Once the stage cools down to room temperature, remove the wafer and inspect the encapsulated graphene moiré superlattice using an optical microscope.

## 4    Definition of Hall bar geometry for graphene moiré superlattice devices

4.1.1 Remove the PC film by immersing the wafer in chloroform for 20 minutes. Rinse the wafer with isopropyl alcohol once taken out from the chloroform solution and blow dry.

4.1.2 Inspect the stack again using an optical microscope to ensure it remains unchanged after the PC film removal. Then, use AFM tapping mode to scan the encapsulated graphene moiré superlattice. Select regions free of bubbles, cracks, and tension as the targeted device area. This process filters out highly disordered regions, improving twist angle consistency across different samples.

4.1.3 Use electron-beam lithography (EBL) to define the contact configuration, then use reactive ion etching (RIE) with a standard hBN etching recipe ($CHF_3$ and $O_2$ mixture) to expose the fresh graphene edges. The Cr/Au one-dimensional edge contacts[29] are thermally evaporated immediately after the etching.

4.1.4 Use EBL and RIE to define the final Hall bar device geometry. This completes the fabrication process of a graphene moiré superlattice device.

**REPRESENTATIVE RESULTS:**
Low temperature transport measurements were performed to characterize the graphene moiré superlattice devices. Starting with MATBG as an example, the typical Landau fan diagram of a high-quality magic-angle device is shown in Fig. 8A (Ref. [22]). The filling factor of the flat bands $v = 4n/n_s$ describes the carrier density of the system, where $n$ is the carrier density and $n_s = 8\theta^2/(\sqrt{3}a^2)$ is the superlattice density ($a$ = 0.246 nm is the graphene lattice constant). Four-fold degenerate Landau fans emanate from the charge neutrality point ($v$ = 0) and superlattice gaps ($v$ = ±4), reflecting the spin and valley degrees of freedom. They break down into two-fold or one-fold degeneracy at higher magnetic fields. Meanwhile, symmetry-broken two-fold degenerate Landau fans emanating from the half-filling ($v$ = ±2) correlated insulating states are also well-identified. The global twist angle can be estimated from the carrier density required to reach the superlattice gaps, which is around 1.06° for this MATBG device. The superconducting dome is distinguishable in this Landau fan diagram, residing underneath the Landau fan around $v$ = -2. The temperature dependent $R_{xx}$ measurement (Ref. [2]; Fig. 8B) at the hole-doped superconducting region of another device with a similar twist angle (1.05°) shows two superconducting domes at both sides of the half-filling state ($v$ = -2). The extracted optimal critical temperature $T_c$ is approximately 1.7 K for this device. In Fig. 8C, the statistics of the optimal doping $T_c$ among the 14 MATBG devices reported in one of the previous works[15] is shown. The peak of $T_c$ locates around 1.08°, which matches with the theoretically predicted magic angle value of TBG[30]. Moreover, $T_c$ at a particular twist angle is also determined by the device quality. The green data points represent two disordered devices (M3 and M8) near the magic angle, showing substantially reduced $T_c$ values. Their temperature dependent $R_{xx}$ linecuts at the optimal doping are given in Fig. 8D, in comparison to the linecuts from high-quality MATBG devices. These results demonstrate the impact of disorder on the device quality, underscoring the necessity of having an optimized fabrication protocol to achieve high-quality graphene moiré superlattice devices.

The optimized fabrication protocol presented in this article has facilitated the investigation of various other graphene-based moiré superlattice structures beyond MATBG, where emergent phenomena are induced by strong correlations and non-trivial topology. In twisted bilayer-bilayer graphene[6], multiple electric displacement field-tunable and spin-polarized correlated insulating states have been observed at integer fillings of the moiré unit cells. Building upon the superconductivity observed in MATBG, a more tunable and strongly coupled superconductor was realized in magic-angle twisted trilayer graphene[10], ultimately leading towards the discovery of robust superconductivity in magic-angle twisted graphene family[13]. On the other hand, superconductivity and strong interaction were also observed in a moiré quasicrystal formed in two-angle twisted trilayer graphene[31]. More recently, intrinsic anomalous Hall effect was uncovered in helical twisted trilayer graphene[32]. Notably, other research groups, employing their own fabrication protocols, had also successfully realized similar graphene moiré superlattices and reported comparable phenomena[5, 7–9, 12, 14].

## DISCUSSION:

Imposing a small twist angle between graphene layers can introduce significant disorder within the heterostructure. Therefore, to maximize the success rate in fabricating high-quality graphene moiré superlattice devices, a highly tunable transfer setup with precise control over position, angle, and temperature is essential. Additionally, several critical aspects during the fabrication process are identified: rigorous flake selection criteria, pre-cleaned bubble-free bottom gates, graphene laser ablation, and, most importantly, smooth pickup of the pre-cut graphene pieces at a submicron speed at room temperature.

The transfer process in this protocol is conducted in a highly tunable, custom-built transfer setup, as shown in Fig. 3A. The entire apparatus is mounted on a passive vibration isolator to minimize vibrational noise. A microscope with 2x, 10x, and 50x long-working-distance objectives is used to visualize the transfer process, with a CMOS camera connected to the computer. The sample stage is based on a direct-drive rotation stage that offers 360° continuous rotation with 2 µrad resolution, featuring a water-cooling system and a central vacuum port to secure the wafer. Additionally, the stage allows movement in the XY directions, controlled by two micro linear actuators. The glass slide with the PC/PDMS stamp can be firmly clamped in a socket (Fig. 3B), with tilting angles adjustable using two goniometers positioned underneath. The socket is mounted on a tunable sliding tray with a manual lock for convenient loading and unloading of the wafer and glass slide. At the bottom of the tray is an XYZ linear stage, of which the Z direction is electronically controlled by an actuator, while XY directions are manually controlled with two vernier micrometers. The transfer process also utilizes two free and open-source software programs: *Inkscape* and *Peek Through*. *Inkscape* is a vector graphics editor used to outline the boundaries of the flake, gate and any debris, serving as a reference for device design and subsequent cleanroom processing. *Peek Through* can make the *Inkscape* window semi-transparent, enabling precise alignment of the flakes in the camera view according to the design drawings in *Inkscape*.

The fabrication of high-quality graphene moiré superlattice devices requires ideal 2D material flakes. Four types of flakes are utilized in the fabrication process: graphite fingers for the electrical gates, graphene as the building block of the moiré superlattice, and hBN for both bottom and top dielectrics. Each type of flake adheres to specific selection criteria. For graphite fingers, a clean, uniform, and elongated rectangular piece with a thickness between 3 nm and 10 nm is preferred (Fig. 1G). Graphite that is too thin may introduce external quantum oscillation signals into the sample layer through capacitive coupling[33], while excessively thick flakes can create additional tension at the edges of the gate. Regarding graphene used for the moiré superlattice, an ideal flake should measure at least 20 µm by 30 µm for each piece after laser cutting. Flakes exhibiting severe inherent tension or cracks need to be avoided. It is preferable that the flake is not surrounded by thick graphite chunks, which can obstruct the wavefront movement and hinder a smooth pickup/dropdown process. An example of an ideal monolayer graphene flake is shown in Fig. 1H, measuring 150 µm in length and 50 µm in width. For hBN, different criteria apply to the bottom and top dielectrics. In both cases, the preferred flake thickness is between 10 nm and 30 nm. Thinner hBN is more fragile, while thicker hBN tends to have a lower breakdown electric field[34]. Additionally, a sufficiently large area with consistent layer thickness is necessary to cover both the graphite gate and the graphene. For hBN used as the top dielectric, it is ideally free of crystal folds or cracks. Most importantly, a straight crystal edge is essential. It functions as an anchor to park the wavefront and selectively pick up the graphene flakes. A representative hBN flake for the top dielectric is shown in Fig. 1I as a reference.

To minimize the complexity involved in fabricating graphene moiré superlattice devices, bubble-free graphite bottom gates are prepared and cleaned ahead of time, as described in the PROTOCOL section 3.2. With the clean gate prepared, the graphene moiré superlattice can be encapsulated on the wafer immediately after pickup, eliminating the need for additional transfer process that can possibly disrupt the twist angle. The essence of bottom gate preparation is to use the hBN to pick up the graphite finger at an elevated temperature, around 80 °C to 100 °C. Ideally, the graphite finger will fully adhere to the hBN flake upon initial contact at one end, spontaneously forming a bubble-free hBN-graphite interface. To achieve a clean gate surface, the bottom gates need to be annealed at high temperature and then tip-cleaned with an AFM in standard contact mode before use. Notably, the tip cleaning process is indispensable, as it cleans away the polymer residue on the gate surface, as shown in Fig. 6C, D.

Conventionally, the "tear & stack" method[35, 36] was used to prepare the graphene moiré superlattices. However, this method relies on the top hBN flake to tear a pristine graphene flake into pieces, thereby generating additional tension. Specifically, the graphene edge defined by this method is not always neat. It sometimes rolls up and develops unexpected folds or cracks. To overcome these difficulties, "laser-cut & stack" method[37, 38] is implemented in this protocol to ablate the graphene. The supercontinuum fiber laser used for graphene laser ablation is integrated into the transfer setup via an additional beam splitter positioned before the camera, as illustrated in Fig. 4A. It features a pulse width of approximately 5 ps and a repetition rate of 40 MHz, with the output filtered to the visible light range (400 nm to 700 nm). During the laser ablation process, the laser is focused into a sub-micron beam spot using a 50x objective. Its light path is confocal with the camera imaging path through the microscope, which enables us to align the laser beam spot to the planned laser-cutting lines and perform the laser ablation process precisely. It is worth noting that an alternative approach based on local anodic oxidation[39] using AFM can also be employed to achieve clean graphene cutting.

With all preconditions optimized, the most critical procedure is smoothly picking up the graphene pieces, aiming to achieve both precise flake alignment and excellent wavefront control. This process is ideally conducted at room temperature to avoid any heat-induced instabilities, such as camera image oscillation caused by air convection, impeded wavefront movement due to increased adhesion between the PC film and the wafer, and twist angle relaxations at elevated temperature. At room temperature, the glass slide engages and disengages at a sub-micron Z-direction speed of 0.02 μm/s, allowing the hBN to slowly pick up the graphene. As the wavefront moves continuously and smoothly during the entire process, bubble formation within the heterostructure is minimized.

In summary, this article presents the detailed workflow of an optimized, experience-informed experimental protocol for fabricating high-quality graphene-based moiré superlattice devices. The transfer and fabrication principles outlined in this protocol can be adapted and extended to construct a wide range of other graphene moiré superlattices, including but not limited to twisted double bilayer graphene[6], helical trilayer graphene[32], and moiré graphene vertical tunnel junction[40]. Future improvements to the transfer setup and fabrication protocol are conceivable. Enclosing the core components of the transfer setup inside a vacuum chamber[41] could significantly reduce surface contamination from the atmosphere, preventing the formation of bubbles that traps air, water, and airborne hydrocarbons[42]. Therefore, cleaner interfaces inside the van der Waals heterostructures with fewer bubbles can be achieved, leading to larger usable device regions in the final stacks. Additionally, replacing the glass slide with PC/PDMS stamps by flexible silicon nitride membranes would enable polymer-free assembly[43], minimizing interlayer contamination resulting from polymer residues. Combining this polymer-free assembly technique with an ultra-high vacuum environment could further eliminate external contaminants, ensuring atomically clean interfaces. Moreover, automated assembly strategies[44, 45] provide a promising avenue for the high-throughput fabrication of graphene moiré superlattice devices, offering unprecedented speed and scalability over the commonly employed artisanal assembly approach, thereby expediting device fabrication. It is anticipated that continued advancements in the fabrication protocol will facilitate the discovery of new emergent phenomena, further enriching the understanding and expanding potential applications of moiré materials.

**ACKNOWLEDGMENTS:**
The authors deeply thank all of the current and former Jarillo-Herrero Group members for developing and optimizing the fabrication protocol. This work has been primarily supported by the Army Research Office MURI W911NF2120147; with support also by the 2DMAGIC MURI FA9550-19-1-0390, the MIT/Microsystems Technology Laboratories Samsung Semiconductor Research Fund, the Sagol WIS-MIT Bridge Program, the National Science Foundation (DMR-1809802), the Gordon and Betty Moore Foundation's EPiQS Initiative through grant GBMF9463, and the Ramon Areces Foundation. This work made use of Harvard's Center for Nanoscale Systems, supported by the NSF (ECS-0335765).

**DISCLOSURES:**
The authors declare no conflict of interest.

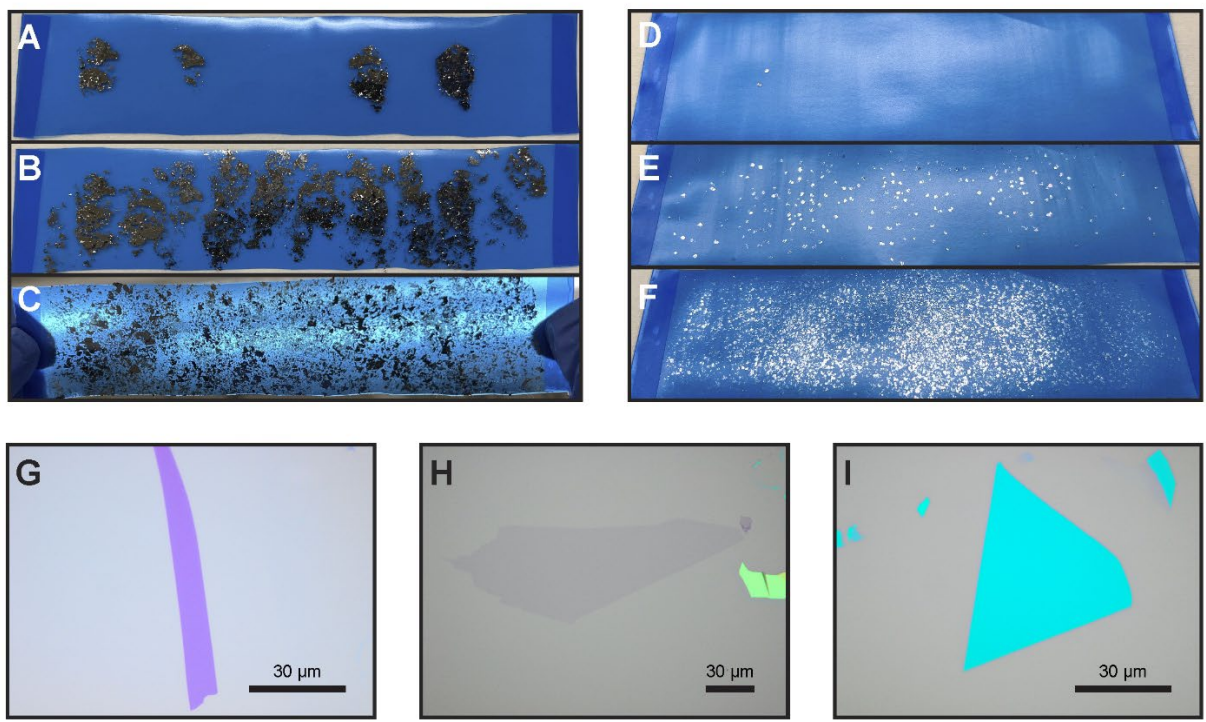

**Fig. 1 Flake exfoliation and selection criteria.** (A-C) Graphene exfoliation. (A) Freshly transferred top-layer graphite crystals onto a new blue tape. (B) Fold and unfold the tape to spread out the graphite crystals on the tape surface. (C) Inspect the tape towards a light source before use. Ideally, most of the area should be covered with semi-transparent graphite islands. (D-F) hBN exfoliation. (D) Select 2 to 3 shiny hBN crystals and put them on the tape. (E) Spread the hBN crystals evenly over the tape by gently folding and unfolding it. (F) Inspect the tape surface using reflected light before use. If most of the crystal surface appears to be colorful, use an additional tape to reduce the crystal thickness. (G-H) Optical microscopic images of the representative 2D material flakes that meet the flake selection criteria, including the graphite finger (G), the monolayer graphene (H), and the top hBN flake (I). (G, I) are captured with a 50x objective, while (H) is captured with a 100x objective. All scale bars denote 30 μm.

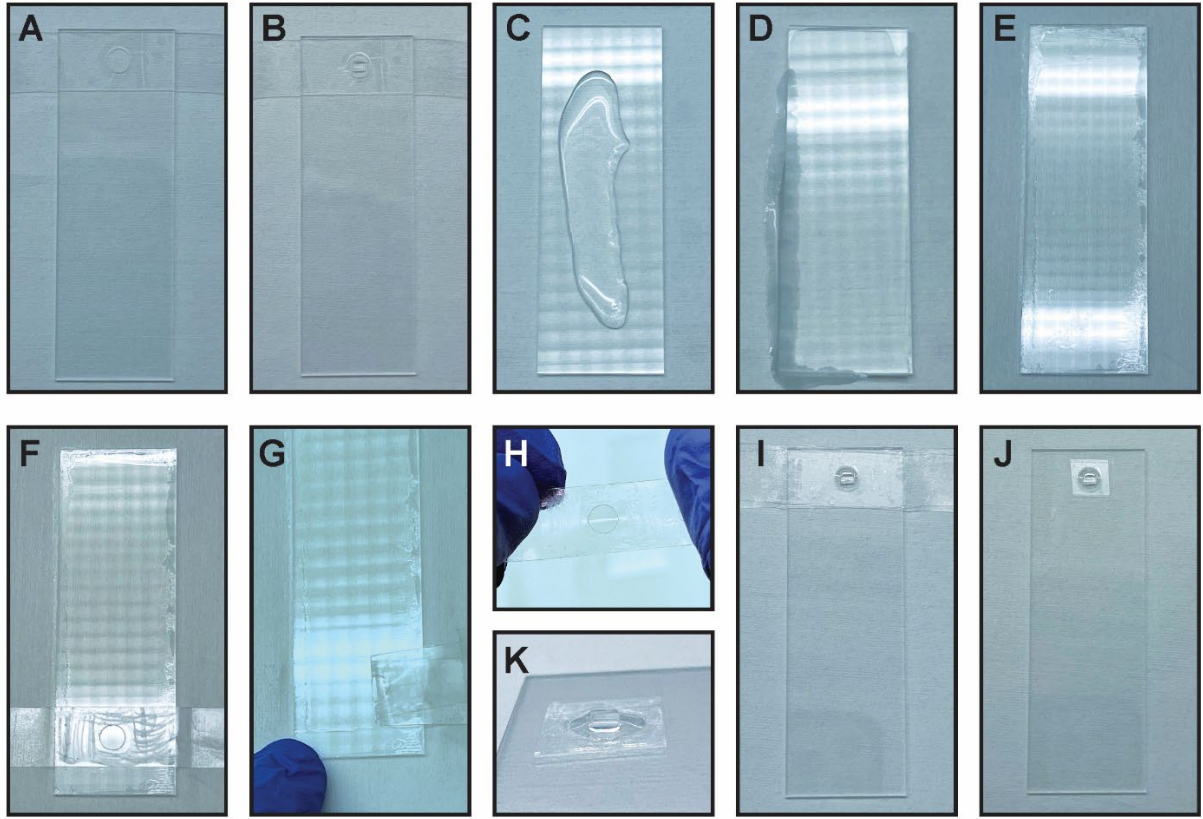

**Fig. 2 Fabrication of glass slides with PC/PDMS stamps.** (A) A double-sided tape with a punched hole was adhered to one end of the glass slide. (B) A clean PDMS stamp was positioned at the center of the punched hole. (C) An appropriate amount of PC in chloroform solution was dispensed onto the glass slide. (D) Another clean glass slide was placed over the bottom one to spread the solution uniformly, and the slides were slid across each other. (E) A uniform PC thin film formed on the bottom slide after the chloroform evaporated. (F) Another double-sided tape with a punched hole was placed over a clean region of the PC film. Scratches around the hole were made by the back of the tweezer to ensure a tight bond between the tape and the PC film. (G) The PC film was cut along the edges of the tape and peeled off gently. (H) A free-standing uniform PC film was visible at the center of the hole. (I) The free-standing PC film was overlaid onto the PDMS stamp by aligning the punched hole in (H) with the one in (B). The double-sided tapes around the hole were pressed firmly together. (J) The additional tape was cut away, leaving only the central PC/PDMS stamp region. (K) Zoom-in view of the PC/PDMS stamp.

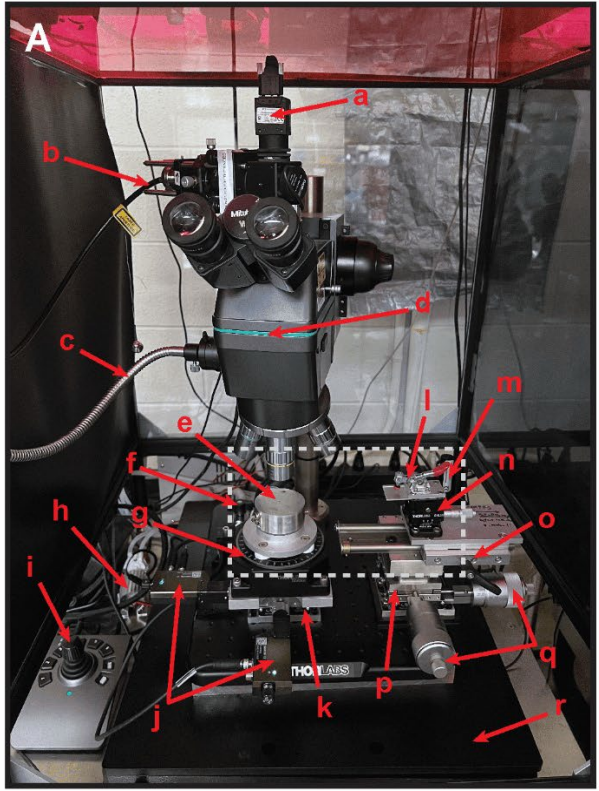
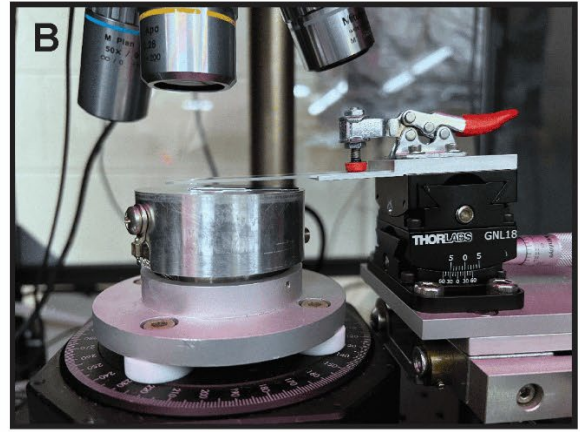

| | |
|---|---|
| a - CMOS camera | j - XY actuators |
| b - fiber laser | k - XY linear stage |
| c - fiber illuminator | l - glass slide socket |
| d - microscope | m - Z actuator |
| e - sample stage | n - goniometers |
| f - vacuum valve | o - sliding tray with lock |
| g - rotation stage | p - XYZ linear stage |
| h - water pump | q - XY vernier micrometers |
| i - XY controllers | r - vibration isolator |

**Fig. 3 Highly tunable custom-built transfer setup.** (A) Front view of the transfer setup, with important components labeled with red arrows and letters. The name of each component is listed on the right. The core region of the setup is enclosed within a grey dashed line box. (B) Zoomed-in view of the core region in (A) at the pre-engage condition, showing that the PC/PDMS stamp is positioned around 2 mm above the Si/SiO$_2$ wafer.

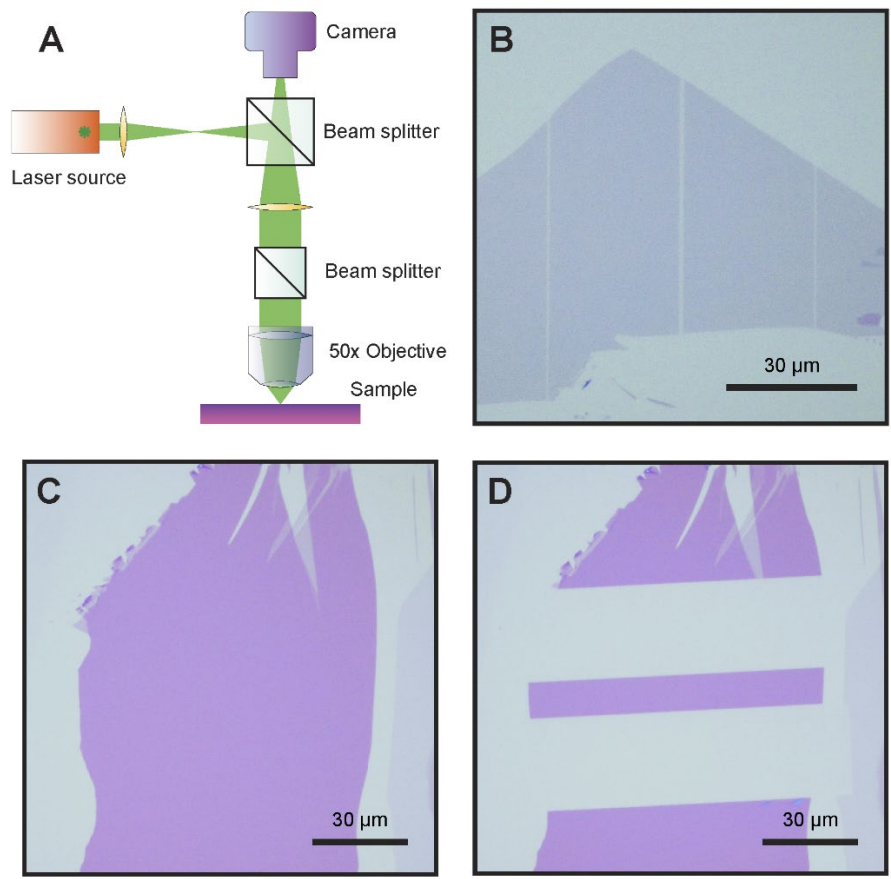

**Fig. 4 Graphene laser ablation.** (A) Schematic of the laser light path in the custom-built transfer setup. The laser beam is confocal with the camera imaging path for precise laser ablation. (B) A monolayer Graphene flake after laser ablation. The middle laser-cutting line is cut twice to increase the gap between adjacent graphene pieces, while the two outer lines separate the core region from the rest of the flake. (C) A graphite flake with uniform thickness but irregular shape. (D) Laser ablation shapes the irregular graphite into a rectangular graphite finger, making it ideal for use as a graphite gate electrode. All scale bars denote 30 μm. Panel A adapted with permission from ref.[38], Springer Nature Ltd.

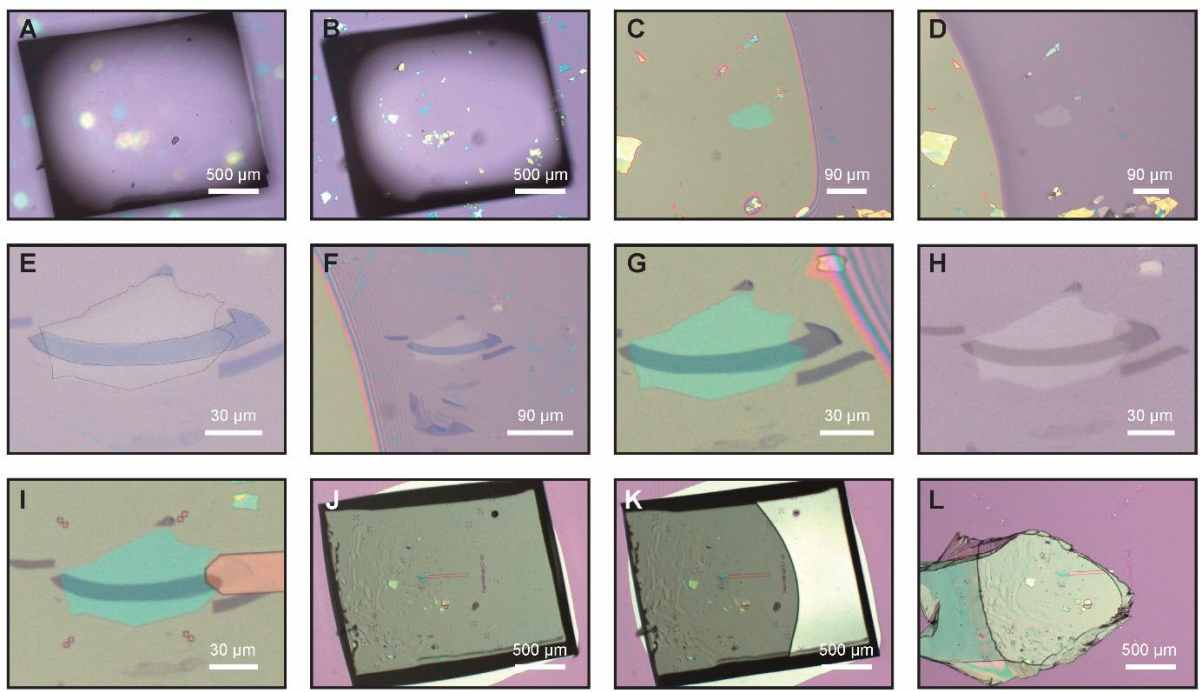

**Fig. 5 Transfer of the bottom graphite gate.** (A) The PC film was focused on, and a clean region was identified under the 2x objective. (B) The hBN flake was focused on and moved to the clean region identified in (A). (C) The stamp contacted the wafer, forming a light green contact region, with the colorful right edge referred to as the wavefront. The PC film fully covered the bottom hBN flake. (D) The bottom hBN flake was picked up, with the flake color appearing semi-transparent. (E) The hBN and the graphite finger were positioned in final alignment before the stamp contacted the wafer. (F) The wavefront approached the graphite finger, where the engaging speed was reduced to 0.5 µm/s. (G) The wavefront completely passed over the graphite finger. (H) The graphite finger was picked up by the bottom hBN flake. (I) The bottom graphite gate was released onto a pre-patterned marker wafer at 160 °C, ensuring contact between the graphite finger and the Ti/PdAu metal finger. (J) The entire PC/PDMS stamp was engaged onto the wafer. (K) The PC film was separated from the PDMS stamp. The transparent wavefront between different colored regions indicated detachment between the PC film and the PDMS stamp. (L) The PC film was torn off from the glass slide, leaving the bottom graphite gate on the wafer. Scale bars in (A,B,J,K,L) denote 500 µm; in (C,D,F) denote 90 µm; and in (E,G,H,I) denote 30 µm.

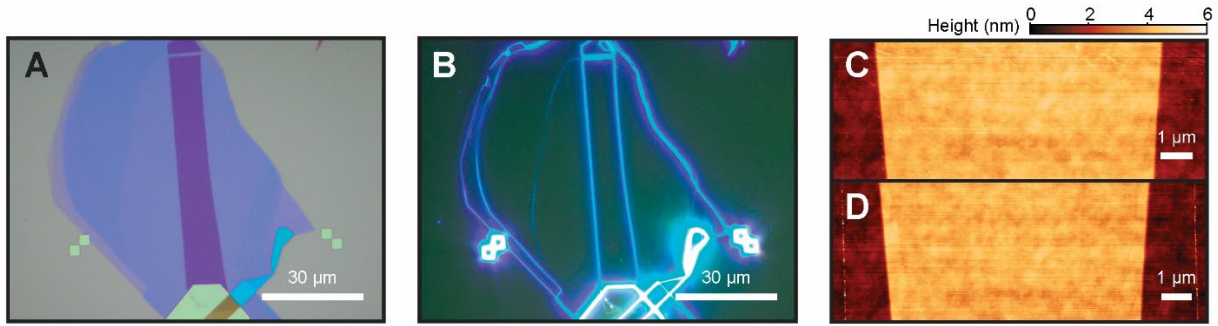

**Fig. 6 Precleaned bubble-free bottom graphite gate.** (A-B) Optical microscopic bright-field (A) and dark-field (B) images of a representative bubble-free bottom graphite gate. AFM topography images of the dirty gate surface before the tip-cleaning process (C) and the clean gate surface after the tip-cleaning process (D). The polymer residue is swept to the side, resulting in two vertical lines as observed in (D). Scale bars in (A, B) denote 30 µm, in (C, D) denote 1 µm.

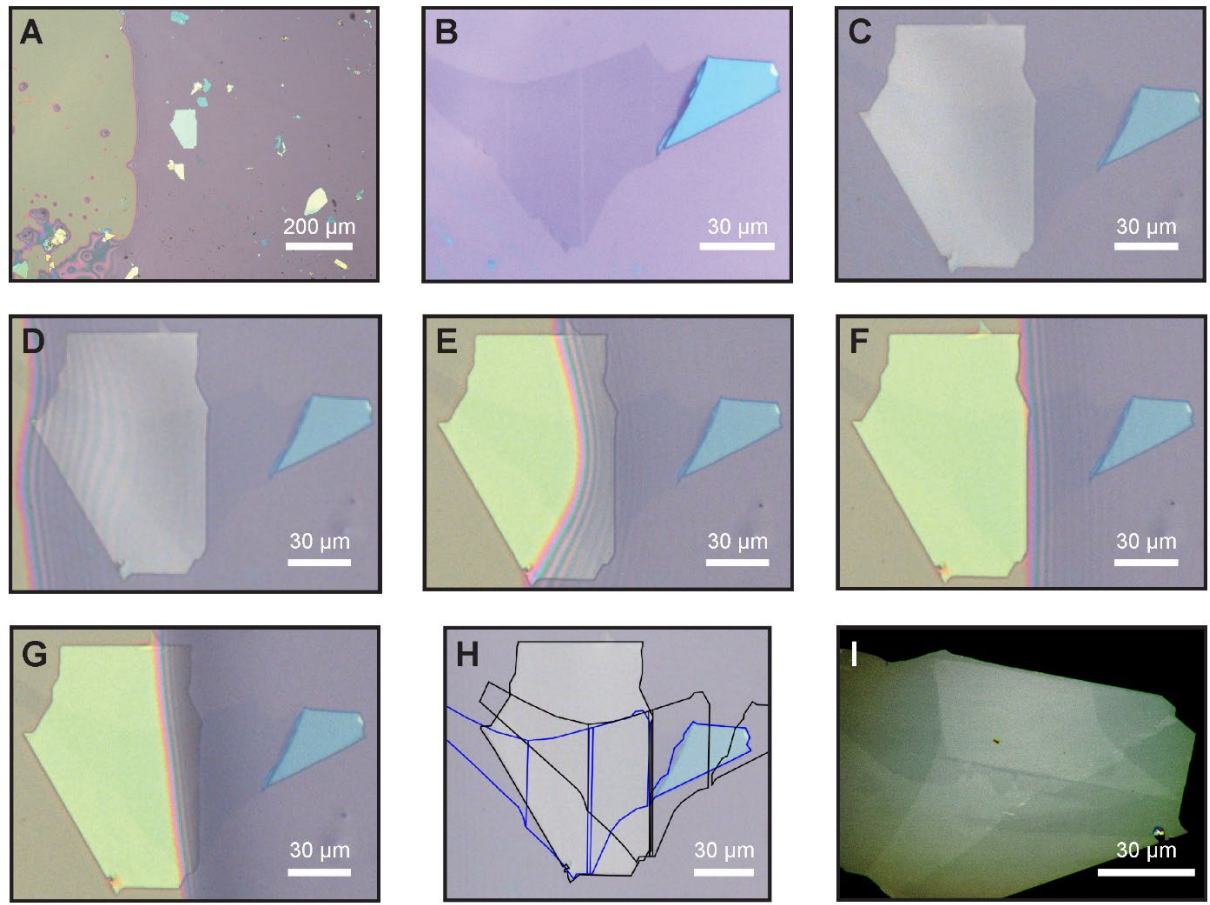

**Fig. 7 Critical steps in the graphene moiré superlattice transfer process.** (A) A zoom-out image taken before picking up the top hBN flake. The contact region between the PC/PDMS stamp and the wafer appears light green. Notably, the wavefront is parallel to the straight edge of the top hBN flake. (B) Graphene flake after laser ablation. The middle laser- cutting line separates the first and the second graphene pieces, which are wider due to double cutting, as mentioned in step 3.1.5. The other two cuts are to isolate the graphene core region from the long graphene tail and thick graphite chunk, which might induce extra strain during the pickup process if not separated. (C) The hBN and graphene are in the final alignment position before picking up the first graphene piece. (D) The wavefront is very close to the hBN left edge, where engaging speed should be reduced to 0.02 μm/s. (E) hBN is slowly coming into contact with the first graphene piece. (F) The wavefront completely passes over the first graphene piece and is pinned by the hBN straight edge. (G) hBN is picking up the graphene smoothly. (H) Flakes are aligned to their respective drawings before picking up the second graphene piece. The boundaries of the hBN and the original graphene drawings are outlined in black, while the copied graphene drawing after rotation is depicted in blue. (I) Optical microscopic image of the top stack fabricated through (A-H), with the color contrast optimally tuned to show the fine features inside the stack. Minimal disorder is identified, indicating a high-quality twisted bilayer graphene top stack. Scale bars in all images represent 30 μm, except in (A), where it represents 200 μm.

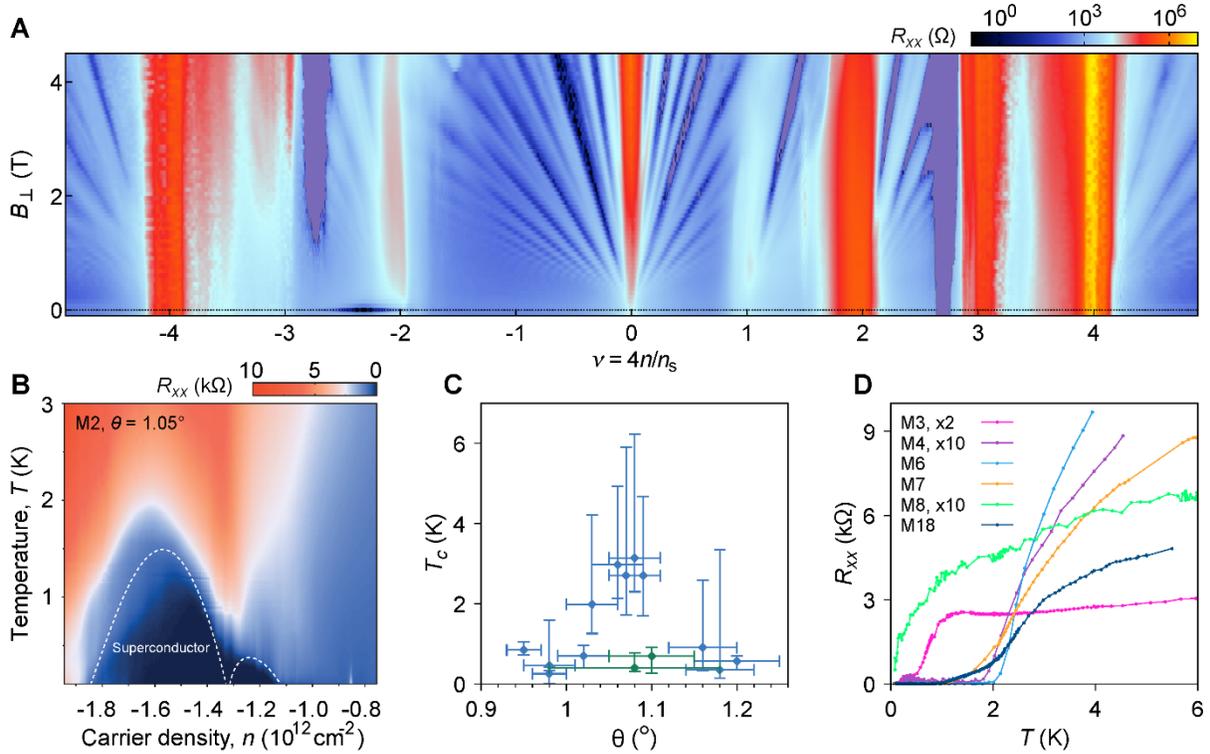

**Fig. 8 Representative measurement results of MATBG.** (A) Landau fan diagram of a high-quality MATBG device with a global twist angle $\theta$ = 1.06°, measured at $T$ = 300 mK. The diagram shows correlated insulating states at integer fillings and hole-side superconductivity. The black dashed line denotes $B_\perp$ = 0 T. (B) Temperature-dependent four-probe resistance $R_{xx}$ measured at the hole-doped superconducting region of another MATBG device with a twist angle $\theta$ = 1.05°. Two superconducting domes are observed next to the $v$ = -2 half-filling state, as indicated by the dashed lines. (C) Statistics of the optimal doping $T_c$ among 14 MATBG devices, determined from 50% of the normal state resistance. It shows a $T_c$ peak around 1.08°. Green data points represent two devices with a substantial level of disorder. Horizontal error bars are estimated during the twist angle extracting process, while the vertical error bars represent $T_c$ determined from 10% and 90% of the normal state resistance. (D) Temperature dependent $R_{xx}$ linecuts measured at optimal doping of two disordered magic-angle devices (M3 and M8) and four high-quality magic-angle devices (M4, M6, M7 and M18). Panels adapted with permission from: A, ref.[22], Springer Nature Ltd; B, ref.[2], Springer Nature Ltd; C, D, ref.[15], AAAS.